\documentclass[english,aps,prd,a4paper,groupedaddress,preprintnumbers,floatfix,
nofootinbib,twocolumn]{revtex4-1}
\usepackage{url}
\usepackage{xcolor}
\usepackage{hyperref}
\hypersetup{colorlinks=true,linkcolor=redLinks,citecolor=greenLinks,
urlcolor=redLinks,
pdfborder={0 0 1}}
\hypersetup{
    bookmarks=true,         
    unicode=false,          
    pdftoolbar=true,        
    pdfmenubar=true,        
    pdffitwindow=false,     
    pdfstartview={FitH},    
    pdftitle={My title},    
    pdfauthor={Author},     
    pdfsubject={Subject},   
    pdfcreator={Creator},   
    pdfproducer={Producer}, 
    pdfkeywords={keyword1, key2, key3}, 
    pdfnewwindow=true,      
    colorlinks=false,       
    linkcolor=red,          
    citecolor=green,        
    filecolor=magenta,      
    urlcolor=cyan           
}
\colorlet{shadecolor}{gray!15}
\definecolor{greenLinks}{rgb}{0, 0.6, 0} 
\definecolor{blueLinks}{rgb}{0, 0, 0.6}
\definecolor{redLinks}{rgb}{0.6, 0, 0}
\definecolor{tempText}{rgb}{0.55, 0.10,0.67}
\definecolor{eprintLinks}{rgb}{0.4, 0.4, 0.4}
\definecolor{journalLinks}{rgb}{0.6, 0, 0}
\newcommand{\MYhref}[3][redLinks]{\href{#2}{\color{#1}{#3}}}%
\usepackage{subfigure}
\usepackage{rotating}
\usepackage{slashed}     

\usepackage[T1]{fontenc} 
\usepackage[utf8]{inputenc}
\usepackage[english]{babel}
\usepackage{amsmath}
\usepackage{graphicx}
\usepackage{color}
\usepackage{mathrsfs}   
\usepackage{amssymb}
\usepackage{hyperref}
\usepackage{dcolumn}
\usepackage{bm}
\usepackage{graphicx}
\def\vev#1{\left\langle #1\right\rangle}

\newcommand {\be} {\begin{equation}}
\newcommand {\ee} {\end{equation}}

\def\vev#1{\left\langle #1\right\rangle}

\usepackage{float}

\def\vev#1{\left\langle #1\right\rangle}

\def\e6{$\mathrm{E(6)}$ }
\def\10{$\mathrm{SO(10)}$ }
\def\21{$\mathrm{SU(2)_L \otimes U(1)_Y}$ }
\def\31{$\mathrm{SU(3)_c \otimes U(1)_Q}$ } 
\def\z4{$\mathrm{Z}_4$}
\def\z2{$\mathrm{Z}_2$}
\def\SM{$\mathrm{SU(3)_c \otimes SU(2)_L \otimes U(1)_Y}$ }
\newcommand{\sm}{{Standard Model }}

\def\3211{$\mathrm{SU(3) \otimes SU(2)_L \otimes U(1)_R \otimes U(1)_{B-L}}$ }
\def\321{$\mathrm{SU(3) \otimes SU(2) \otimes U(1)}$ }
\def\422{$\mathrm{SU(4) \otimes SU(2) \otimes SU(2)_R}$ }

\begin{document}

\title{A flavour physics scenario for the $750$ GeV diphoton anomaly}

\author{Cesar Bonilla}
\email{cesar.bonilla@ific.uv.es}
\affiliation{Institut de F\'{i}sica Corpuscular --
  C.S.I.C./Universitat de Val\`{e}ncia, Parc Cient\'ific de Paterna.\\
 C/ Catedr\'atico Jos\'e Beltr\'an, 2 E-46980 Paterna (Valencia) - SPAIN}

\author{Miguel Nebot}
\email{nebot@ific.uv.es}
\affiliation{Institut de F\'{i}sica Corpuscular --
  C.S.I.C./Universitat de Val\`{e}ncia, Parc Cient\'ific de Paterna.\\
 C/ Catedr\'atico Jos\'e Beltr\'an, 2 E-46980 Paterna (Valencia) - SPAIN}

 \author{Rahul Srivastava}
\email{rahuls@imsc.res.in}
\affiliation{The Institute of Mathematical Sciences, \\
IV Cross Road, CIT Campus, Taramani,
Chennai 600 113, India}

 \author{\ Jos\'e W. F. Valle }
   \email{valle@ific.uv.es}
 \affiliation{Institut de F\'{i}sica Corpuscular --
  C.S.I.C./Universitat de Val\`{e}ncia, Parc Cient\'ific de Paterna.\\
 C/ Catedr\'atico Jos\'e Beltr\'an, 2 E-46980 Paterna (Valencia) - SPAIN}


\begin{abstract}

  A simple variant of a realistic flavour symmetry scheme for fermion
  masses and mixings provides a possible interpretation of the
  diphoton anomaly as an electroweak singlet ``flavon''. The existence
  of TeV scale vector-like T-quarks required to provide adequate
  values for CKM parameters can also naturally account for the
  diphoton anomaly. Correlations between $V_{ub}$ and $V_{cb}$ with
  the vector-like T-quark mass can be predicted. Should the diphoton
  anomaly survive in a future Run, our proposed interpretation can
  also be tested in upcoming B and LHC studies.

\end{abstract}

\maketitle




The ATLAS~\cite{atlas750} and CMS~\cite{cms750} collaborations
have presented first results obtained from proton collisions at the
LHC with 13 TeV center-of-mass energy.
The ATLAS collaboration sees a bump in the invariant mass distribution
of diphoton events at 750 GeV, with a 3.9 sigma significance, while
CMS sees a 2.6 sigma excess at roughly the same value. Taking these hints
at face value, we suggest a possible theoretical framework to interpret
these findings.
We propose that the new particle is a \SM singlet scalar boson
carrying a flavour quantum number. Our proposed framework accounts for
three important aspects of the flavor puzzle:
\begin{itemize}
\item the observed value of the Cabbibo angle arises mainly from
  the down-type quark sector through the Gatto-Sartori-Tonin
  relation~\cite{Gatto:1968ss};
\item the observed pattern of neutrino
  oscillations~\cite{Forero:2014bxa} is reproduced in a restricted
  parameter range~\cite{Morisi:2013eca};
\item the observed values of the ``down-type'' fermions is
  well-described by the generalized b-tau unification
  formula~\cite{Morisi:2011pt,Morisi:2013eca,King:2013hj,Bonilla:2014xla}
\begin{equation}\label{eq:massrelation}
\frac{m_{\tau}}{\sqrt{m_{e}m_{\mu}}}\approx \frac{m_{b}}{\sqrt{m_{s}m_{d}}}, 
\end{equation}
predicted by the flavour symmetry of the model.
\end{itemize}
There are in principle several possible realizations of the 750 GeV
anomaly as a flavon~\cite{Morisi:2012fg,King:2014nza} : a
flavor--carrying \SM singlet scalar. Our main idea is to obtain a
scheme where the CERN anomaly may be probed also in the flavor
sector. For this purpose we consider a simple variant of that proposed
in~\cite{Morisi:2013eca} in order to address the points above.
Phenomenological consistency of the model requires the presence of
vector--like fermions in order to account for the observations in the
quark sector. Their presence can naturally account for a production
cross section of the scalar anomaly through gluon--gluon fusion
similar to that indicated by ATLAS and CMS~\cite{Morisi:2013eca}
\footnote{Indeed vector-like fermions have been suggested to account
  for the diphoton anomaly. For an extensive reference set
  see~\cite{Staub:2016dxq}.}

Here we investigate the allowed parameter space of our scheme which
provides an adequate joint description of CKM physics describing the B
sector and the recent CERN diphoton data, illustrating how the two
aspects are inter-related in our scheme. For definiteness and
simplicity, here we focus on a nonsupersymmetric version of the model
discussed in \cite{Morisi:2013eca}.  The charge assignments for the
fields is as shown in Table \ref{tab1}
\begin{table}[H]
\begin{center}
\begin{tabular}{| c | c c c c c|c c|| c c|| c c || c |}
\hline
 Fields & $L$ & $E^c$ & $Q$  & $U^c$ & $D^c$ & $H^u$ & $ H^d$ & $T$ & $T^c$ & 
$\sigma$ & $\sigma'$ & $ \xi $ \\
\hline
\hline
$\mathrm{SU(2)_L}$  & $2$ & $1$ & $2$ & $1$ & $1$ & $2$   & $2$  & $1$ & $1$ & 
$1$ & $1$  & $1$  \\
\hline
$A_4$ & $3$ & $3$ & $3$ & $3$ & $3$ & $3$ & $3$ & $1$ & $1$ & $3$ & $3$ & $1$   
\\
\hline
$\mathrm{Z}_4$  & $1$ & $1$ & $1$ & $1$ & $1$ & $1$ & $1$ & $\omega$ & $\omega^2$ 
& $\omega^3$ & $\omega^2$  & $\omega$  \\ 
\hline
\end{tabular}\caption{Matter content of the model, where $\omega^4=1$.} 
\label{tab1}
\end{center}
\end{table}

Here, $T, T^c$ are a pair of vector like ``quarks'' transforming as
$(3, 1, 4/3)$ and $(\bar{3}, 1, -4/3)$ under the \sm gauge group
\SM. The scalars $\sigma, \sigma'$ are singlets under \SM but
transform as $A_4$ triplets and carry $\mathrm{Z}_4$ charge. The
scalar $\xi$ is also a singlet under \sm as well as under the $A_4$
symmetry but transforms as $\omega$ under the $\mathrm{Z}_4$ symmetry.
In addition to the above charges, the scalars and fermions also carry
an additional $\mathrm{Z}_2$ charge such that the scalar $H^u$ only
couples to the up-type quarks, while $H^d$ only couples to the
down-type quarks and charged leptons (this $\mathrm{Z}_2$ symmetry
would not be needed if supersymmetry were assumed). The invariant
Yukawa Lagrangian of the model is given by,
\begin{eqnarray}
 \mathcal{L}_f & = & y^u_{ijk} Q_i H^u_j U^c_k + y^d_{ijk} Q_i H^d_j D^c_k + 
y^l_{ijk} L_i H^d_j E^c_k \nonumber \\
& + & X' T U^c_i \sigma_i + \frac{Y'}{\Lambda} Q_i (H^u \cdot \sigma')_i T^c + 
y_T T T^c  \xi 
\label{yuk}
\end{eqnarray}
where we take all couplings $y_T$, $y^a_{ijk}$ and $X$, $Y$ as real
for simplicity; $a = u,d,l$ and $i,j,k = 1,2,3$.

{}Following Ref.~\cite{Morisi:2013eca} after electroweak symmetry
breaking and requiring certain hierarchy in the flavon vevs,
$\vev{H^{u,d}} = (v^{u,d},\varepsilon^{u,d}_1,\varepsilon^{u,d}_2)$
where $\varepsilon_{1,2}^u \ll v^u$ and $\varepsilon_{1,2}^d\ll v^d$,
one gets the mass relation between the down-type quarks and charged
leptons given by Eq.~(\ref{eq:massrelation}).
The up-type quark sector gets modified due to the presence of vector
like quarks so that the full up-type quark mass matrix is $4 \times 4$
and given by
\begin{eqnarray}
 \label{Mu}
M_{u} = \left( \begin{array}{cccc} 
0                & a^u \alpha^u          & b^u           & Y'_1          \\
b^u \alpha^u     & 0                     &  a^u r^u      & Y'_2          \\
a^u              & b^u r^u               & 0             & Y'_3         \\
X'_1             & X'_2                  & X'_3          &   M'_T        
                      \\ 
\end{array}\right) 
\end{eqnarray}
where $a^u = y_1^u \varepsilon_1^u $, $b^u = y_2^u \varepsilon_1^u$;
$y_{1,2}^u$ being the only two possible Yukawa couplings arising from
the $A_4$-tensor in Eq.~(\ref{yuk}). Also, $r^u = v^u/\varepsilon_1^u$
and $\alpha^u = \varepsilon_2^u / \varepsilon_1^u$. Moreover, $X'_i =
X' \vev{\sigma_i}$, $Y'_i= Y' \vev{(H^u\cdot \sigma')_i}/\Lambda$ 
and $M'_T = y_T \vev{\xi}$. The mass matrix in Eq.~\ref{Mu} is the 
same as that obtained in \cite{Morisi:2013eca} where the detailed 
treatment of the Yukawa sector is given.
Notice that the addition of vector quarks only changes the up sector
mass matrix, the down sector mass matrix remaining unchanged and thus
the relation in Eq.~(\ref{eq:massrelation}) remains unchanged, see
\cite{Morisi:2013eca} for further details.

The scalar sector of the model consists of $\mathrm{SU(2)_L}$ doublet
scalars $H^u, H^d$ both transforming as triplets under the $A_4$
symmetry. In addition it contains three types of $\mathrm{SU(2)_L}$
singlet scalars with $\sigma, \sigma' \sim 3$ under $A_4$ while $\xi
\sim 1$ under the $A_4$ symmetry. In order to illustrate how our
candidate scalar can account for the $750$ GeV di-photon excess, we
consider a simplified scenario.
  Neglecting the mixing between the $\mathrm{SU(2)_L}$ doublet and
  singlet scalars  $s^{0}=(\xi^0,\sigma^0,\sigma^{\prime0})$ we can
  phenomenologically express the various scalar mass eigenstates as
  follows
\begin{eqnarray}
   h_i & = &  \mathcal{U}_{ij} H^{0}_{j},\,\, \ \ (i,j=1,...,6) \nonumber \\
   \chi_m & = & \mathcal{O}_{mn} s^{0}_{n}, \ \ (m,n=1,...,7) 
 \label{pscalarmass}
\end{eqnarray}
Under this approximation the singlet scalars $\chi_i$ can be further
decomposed as
 \begin{eqnarray}
   \zeta \equiv \chi_1 &=&  \mathcal{O}_{11} \xi^{0}+ \mathcal{O}_{1n} s^{0}_n, \\
    \chi_m &=& \mathcal{O}_{mn} s^{0}_{n}, \nonumber 
  \label{sscalarmass}
 \end{eqnarray}
 with $m,n=2,...,7$ and we have identified the flavon field mass
 eigenstate $\zeta$ as our $750$ GeV resonance candidate. Then, this
 flavon field is composed predominantly of $\mathrm{SU(2)_L}$ singlet
 scalars.  Note that the rotation matrix $\mathcal{O}$ determines the
 mixing amongst the singlet scalars that form the two $A_4$-triplets
 and the $A_4$-singlet $\xi$.

 At the LHC $\zeta$ will be predominately produced through gluon-gluon
 fusion via a triangle loop involving the vector like T-quarks. In the
 absence of mixing between the $\mathrm{SU(2)_L}$ doublet and singlet
 scalars the tree level coupling of $\zeta$ to $W, Z$ bosons can be
 neglected.
 Similarly, the coupling of $\zeta$ to down-type fermions is also
 negligible.  However the coupling of $\zeta$ to up-type quarks is
 determined by the off-diagonal elements of Eq.~(\ref{Mu}). Thus,
 $\zeta$ predominantly couples to the vector-like quarks $T, T^c$. As
 we show below, the flavour constraints require the vector-like quarks
 to be quite heavy so that, for a large range of parameters we have
 $m_T > m_\zeta/2$. Thus, the decay of $\zeta$ to $T, T^c$ is also
 kinematically forbidden.

\begin{center}
\begin{figure}[h!]
\subfigure[\label{Fig:Decays:gaga}]{\includegraphics[width=0.2\textwidth]{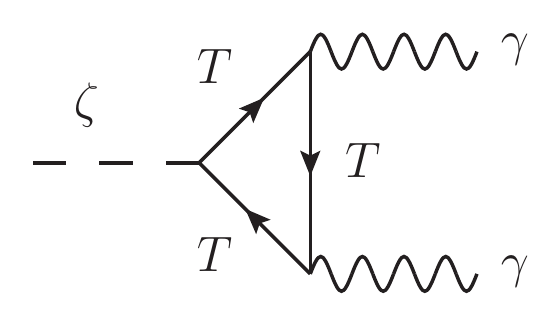}}\quad 
\subfigure[\label{Fig:Decays:glgl}]{\includegraphics[width=0.2\textwidth]{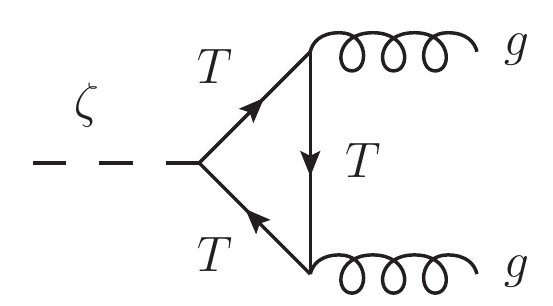}}
\caption{Decays of the 750 GeV scalar to gluons and
  photons.\label{Fig:Decays}}
\end{figure}
\end{center}

 Therefore, $\zeta$ predominantly decays to photons and gluons through
 the triangle loop involving $T, T^c$, as shown in
 Fig.~(\ref{Fig:Decays}) and to up-type quarks through tree-level
 mixing.  \vspace{-0.5cm}
  \vspace{-0.9cm}
Apart from the above channels, $\zeta$ can also decay to $Z \gamma$
and $ZZ$ through analogous triangle loops involving $T, T^c$.
Since $m^2_Z << m^2_\zeta$ the decay widths to $\gamma \gamma$, $Z
\gamma$ and $Z Z$ channels are proportional to each other, so that,
 \begin{eqnarray}
 \frac{\Gamma_{Z \gamma}}{\Gamma_{\gamma \gamma}} \approx  2 \, \tan^2 
\theta_W \ \ \text{and}\ \
 \frac{\Gamma_{Z Z}}{\Gamma_{\gamma \gamma}}  \approx   \tan^4 \theta_W
 \end{eqnarray} 
 where $\theta_W$ is the weak mixing angle. In general $\zeta$ can
 also decay to two Higgs scalars as we discuss below.

 Thus $\zeta$ seems, indeed, an ideal candidate to explain the 750 GeV
 di-photon excess recently observed at the LHC.  Both $g$ and $\gamma$
 couple to $T, T^c$ through gauge interactions with interaction
 strength proportional to $\alpha_s, \alpha$, the strong and
 electromagnetic coupling constants respectively. One can write down
 the effective Lagrangian for the coupling of $\zeta$ with gluons, $Z$
 boson and photons which is
\begin{eqnarray}
  \mathcal{L}_{\rm{eff}} & = & \frac{ c_\gamma}{4} \, \zeta \, F^{\mu \nu} 
F_{\mu \nu} + \frac{c_g}{4}  \, \zeta \, G^{\mu \nu} G_{\mu \nu} + \frac{ 
c_{Z\gamma}}{2} \, \zeta \, F^{\mu \nu} Z_{\mu \nu} \nonumber \\
  & + & \frac{c_{ZZ}}{4}  \, \zeta \, Z^{\mu \nu} Z_{\mu \nu}
 \label{effl}
\end{eqnarray} 
where $F^{\mu \nu}, G^{\mu \nu}$ are the usual electromagnetic and
colour field strength tensors and $Z_{\mu \nu}$ is the field strength
tensor for the $Z$ boson.

\begin{figure}[!h] 
\includegraphics[width= 0.4\textwidth]{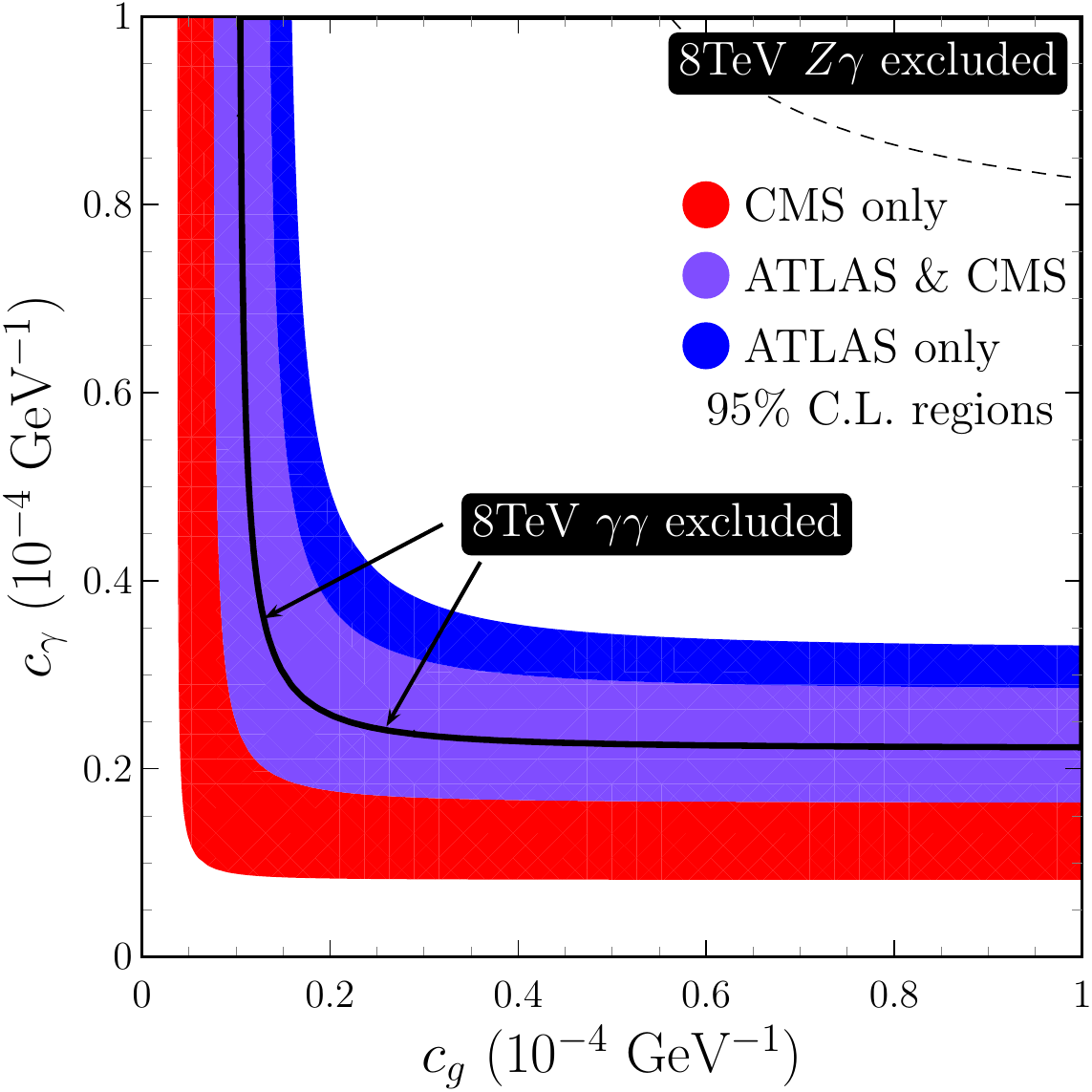}
\caption{\label{fig7} Ranges of effective couplings $c_\gamma$ and
  $c_g$ required to explain the diphoton excess at 95\% confidence
  level. Also shown are the regions excluded by 8 TeV data.} 
\end{figure}

\begin{figure}[H]
\includegraphics[width= 0.4\textwidth]{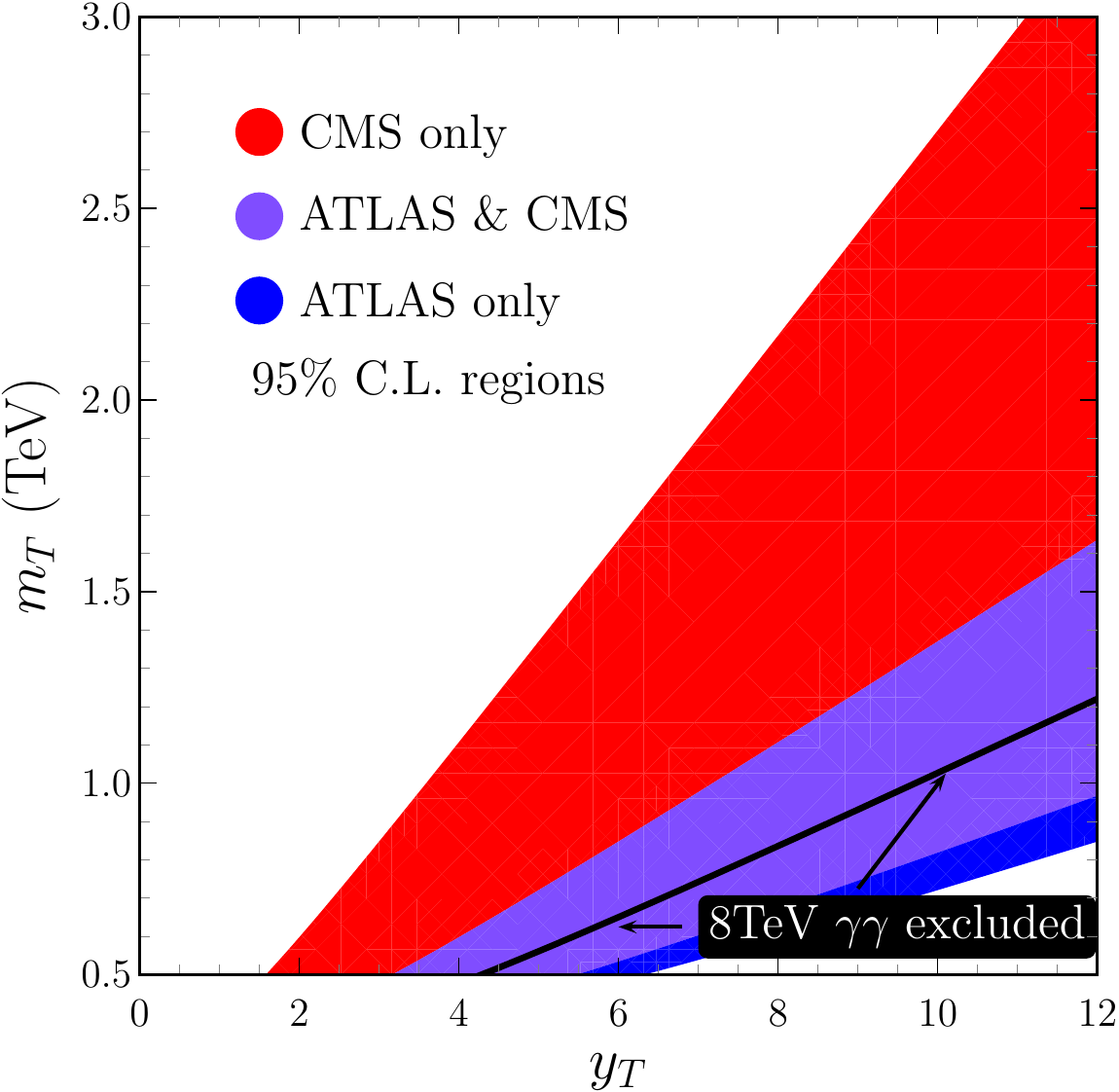}
\caption{\label{fig8} Required vector-like quark mass $m_T$ and Yukawa
  coupling $y_T$ to explain the 750 GeV excess. Also seen are the
  regions excluded by 8 TeV data.}
\end{figure}

In Fig.~\ref{fig7} we show the allowed ranges for the effective
couplings required to account for the 750 GeV di-photon excess for
both the CMS and ATLAS experiments within $95 \%$ confidence level. In
the 8 TeV run neither ATLAS nor CMS have seen any statistically
significant excess in any of the $\gamma \gamma$, $Z \gamma$ and $ZZ$
channels. The constraints from 8 TeV run on production times branching
fraction $\sigma \times Br(\zeta \to ff)$; $f = g,Z,\gamma$ for these
decay channels can be obtained
from~\cite{Aad:2014fha,Aad:2015kna,Aad:2015mna,CMS-PAS-HIG-14-006}.
In Fig.~\ref{fig7}, we have also included the constraints from the
non-observation of any signal excess in the $\gamma \gamma$, $Z
\gamma$ and $ZZ$ in the 8 TeV run. The upper colored region is
consistent only with ATLAS data, while the lower one is consistent with
only CMS data, while the middle region is consistent with both CMS
and ATLAS. The solid and dashed lines delimit the regions disallowed
by 8 TeV data for $\zeta \to \gamma \gamma$ decay and $\zeta \to Z
\gamma$ decay, respectively. The constraints from $\zeta \to g g$ as
well as $\zeta \to Z Z$ decays are rather weak and are not shown in
the graph.
The value of the effective couplings $c_\gamma$ and $c_g$ are
determined by only two parameters, namely the mass $m_T$ of the vector
quarks and the strength of the Yukawa coupling $\zeta T T^c$. The
allowed parameter ranges for the mass $m_T$ and Yukawa coupling $y_T$
for both CMS and ATLAS experiments as well as the constraints from the
8 TeV run are shown in Fig.~\ref{fig8}.  In plotting Fig.~\ref{fig8}
we have required that all the Yukawa couplings remain perturbative
over the entire range of parameter space. The color scheme of
Fig.~\ref{fig8} is the same as that of Fig.~\ref{fig7}.

As shown in Fig.~\ref{fig8}, the decay $\zeta \to \gamma \gamma$ in
our model can explain the diphoton excess observed by both CMS and
ATLAS experiments. Although the non-observation of similar diphoton
excess in the 8 TeV run puts severe constraints on the allowed
parameter range, our model still has enough freedom to reconcile these
restrictions with the observed 13 TeV excess.
\begin{figure}[h!]
\begin{center}
  \subfigure[$|V_{ub}|$ vs. $m_T$; 68\%, 95\% and 99\% C.L. regions
  (darker to lighter) are
  represented.\label{fig:Correlations:Aa}]{\includegraphics[width=0.4\textwidth]{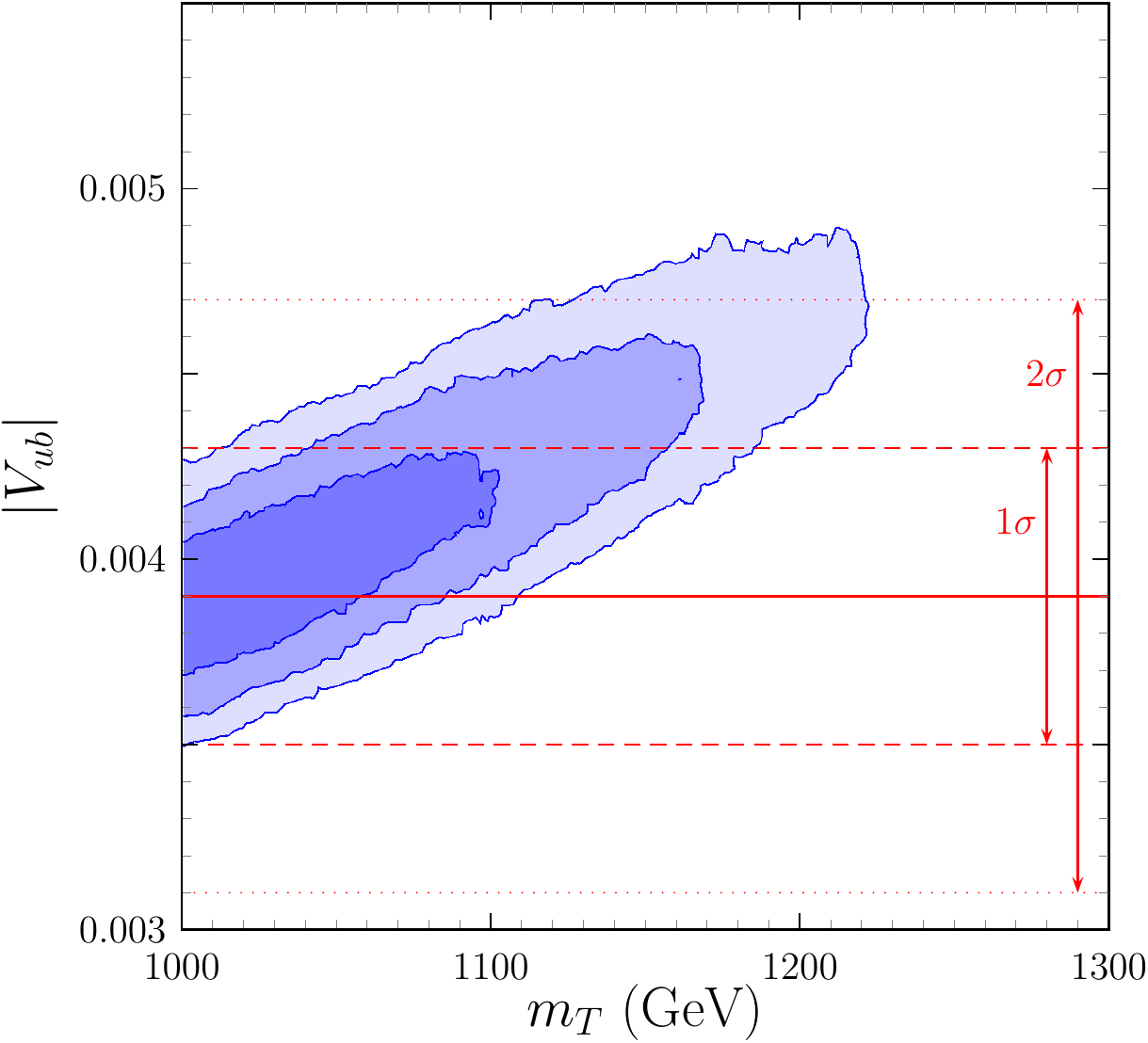}}\qquad
  \subfigure[$|V_{cb}|$
  vs. $|V_{ub}|$\label{fig:Correlations:Ab}]{\includegraphics[width=0.4\textwidth]{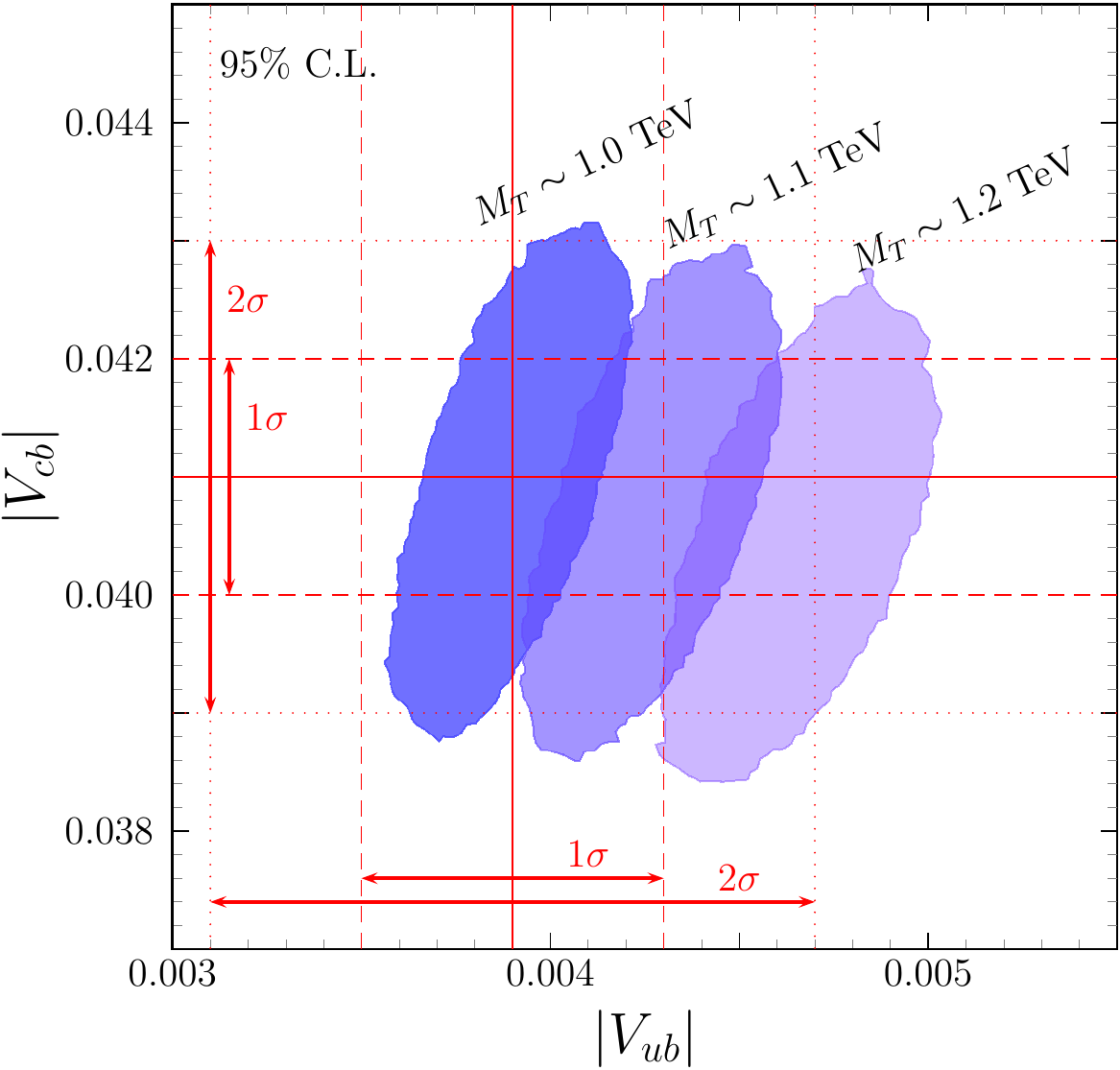}}
\caption{Scenario (A), free $\{X_2,Y_3,M_{44}\}$; experimental central
  values and $1\sigma$, $2\sigma$ intervals are shown for $|V_{ub}|$
  and $|V_{cb}|$\label{fig:Correlations:A}.}
\end{center}
\end{figure}
\begin{figure}[t!]
\begin{center}
  \subfigure[$|V_{ub}|$
  vs. $m_T$\label{fig:Correlations:Ba}]{\includegraphics[width=0.4\textwidth]{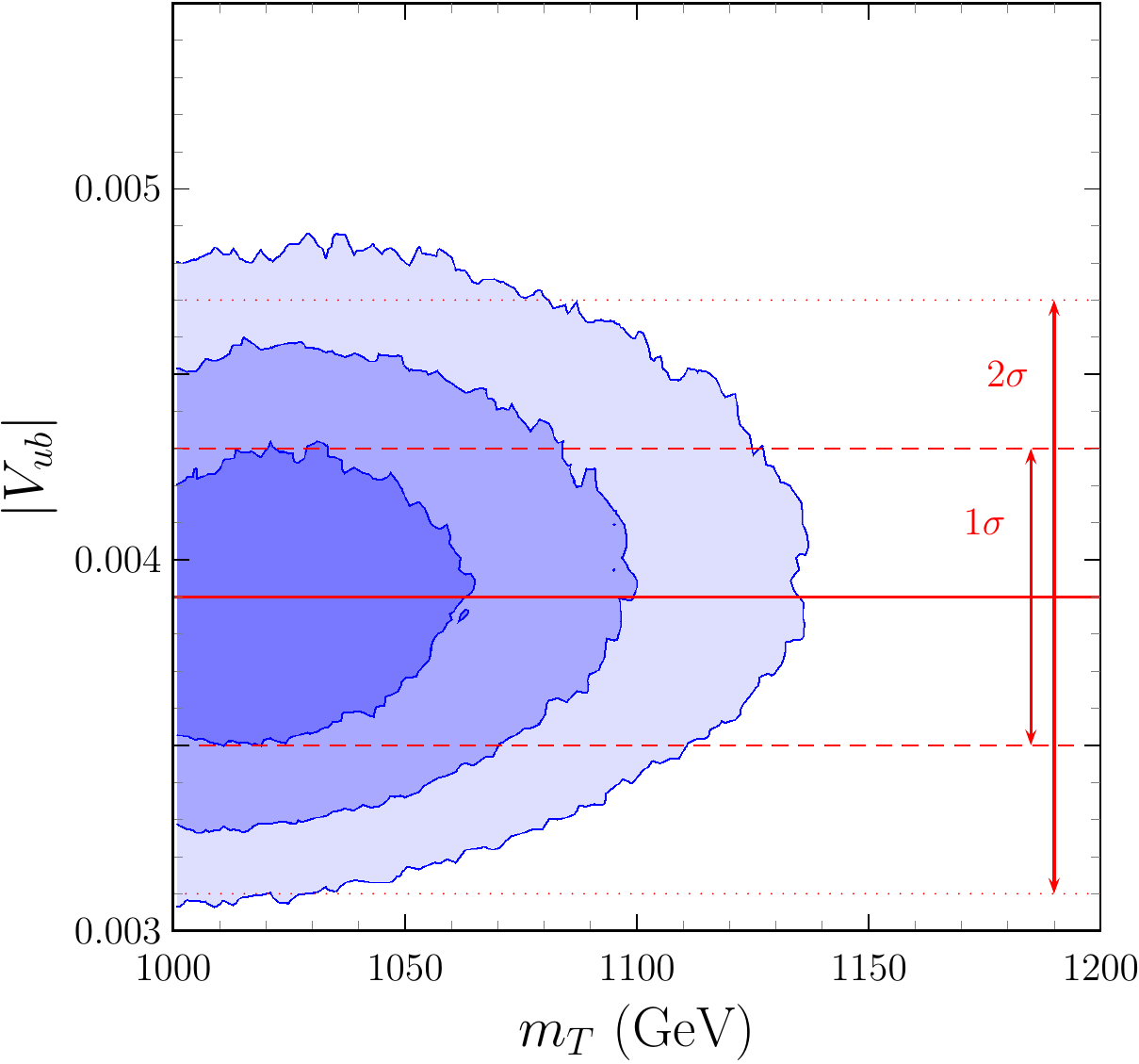}}\qquad
  \subfigure[$|V_{cb}|$
  vs. $m_T$\label{fig:Correlations:Bb}]{\includegraphics[width=0.4\textwidth]{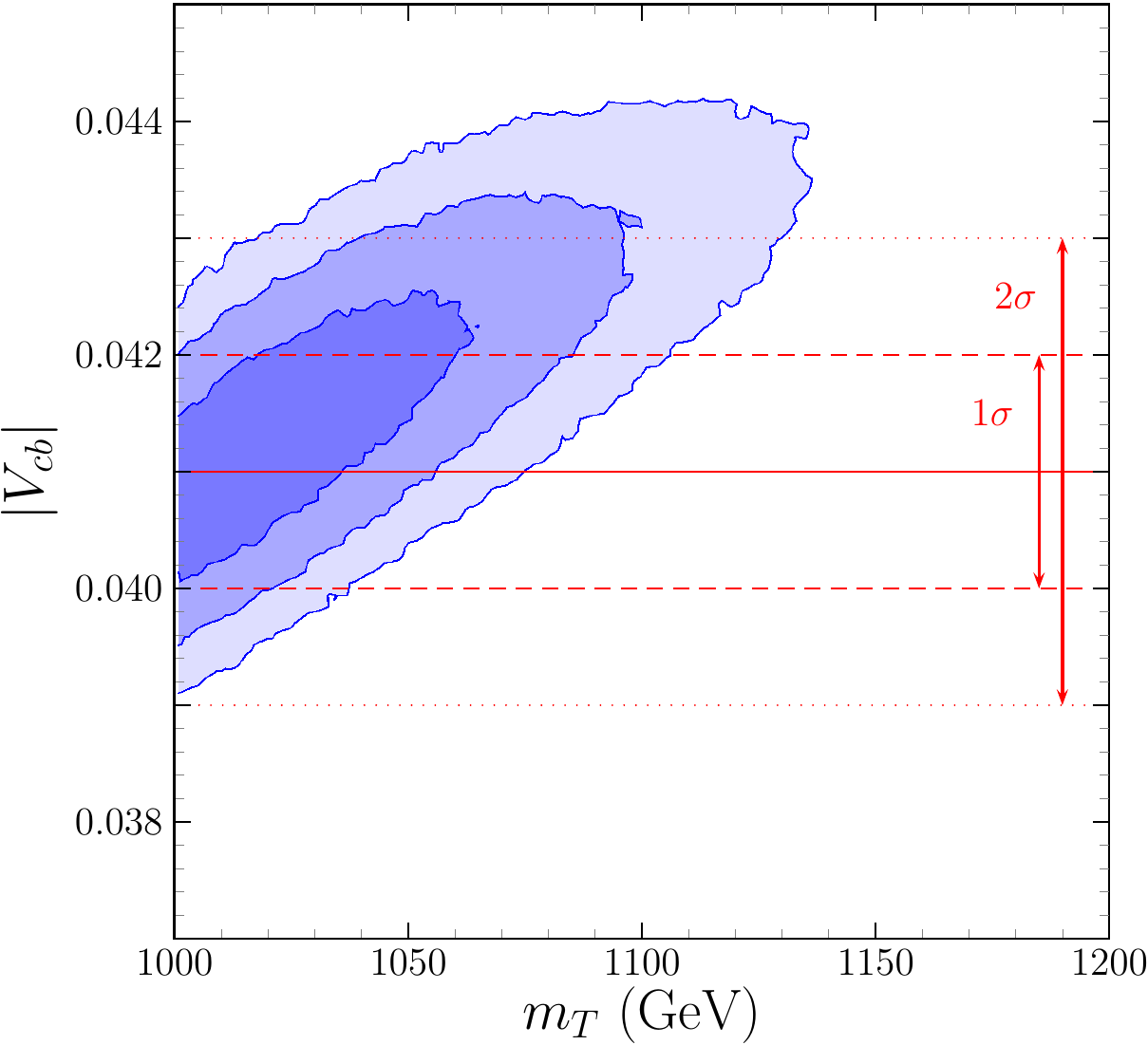}}
\caption{Scenario (B), free $X_i$; 68\%, 95\% and 99\% C.L. regions
  (darker to lighter) are represented; experimental central values and
  $1\sigma$, $2\sigma$ intervals are shown for $|V_{ub}|$ and
  $|V_{cb}|$\label{fig:Correlations:B}.}
\end{center}
\end{figure}
A key feature of our proposal is the identification of the 750 GeV
anomaly as a flavon, i.e. a scalar state that carries flavour
information. Indeed, our scalar $\zeta$ is directly coupled to the
quarks, and hence to flavour, so we expect potential correlations
between CKM physics and the properties of the observed anomaly.
Therefore, in addition to the the quark and lepton masses, related by
Eq.~(\ref{eq:massrelation}), the measured neutrino oscillation
parameters~\cite{Forero:2014bxa}, there are restrictions on the model
parameters that come from the consistency of the quark sector, such as the
measured quark mixing parameters~\cite{Agashe:2014kda}.
In order to explore these implications of the model, one must check,
as in Ref. \cite{Morisi:2013eca} or other generic vector-like
scenarios \cite{Botella:2008qm,Botella:2012ju}, that ours can
indeed adequately reproduce CKM physics. To do this we include a
selected set of additional observables sensitive to the new
vector-like quark $T$ and to the deviations of the CKM matrix from the
standard $3\times 3$ unitary form. In particular, we include:
\begin{enumerate}
\item neutral meson mixing constraints: in $B_d^0$ -- $\bar B_d^0$ and
  $B_s^0$ -- $\bar B_s^0$ systems, mass differences and ``golden'' CP
  asymmetries in $B_d\to J/\Psi K_S$, $B_s\to J/\Psi\Phi$ decays,
  bounds on the short distance contribution to the mass difference in
  the $D^0$ -- $\bar D^0$ system, and indirect and direct CP violation
  parameters $\epsilon_K$ and $\epsilon^\prime/\epsilon_K$ for the
  $K^0$ -- $\bar K^0$ system;
\item rare decays induced by different quark level transitions: 
$B_s\to\mu^+\mu^-$, $B_d\to\mu^+\mu^-$, $B\to X_s\gamma$, $K_L\to 
\pi^0\nu\bar\nu$, $K^+\to \pi^+\nu\bar\nu$, and short distance contributions to 
$D^0\to\mu^+\mu^-$ and $K_L\to\mu^+\mu^-$.
\end{enumerate}
Furthermore, to reflect LHC bounds on direct production of the new
vector-like quark, we restrict our analysis to values of the mass
$m_T>1$ TeV~\cite{Aad:2016qpo}.
Once compliance with this set of constraints is ensured, we can
address, in addition to the 750 GeV diphoton hint, other features of
the model, in particular flavour related ones, like correlations among
different observables. Due to the complexity of the problem in terms
of the number of independent parameters, we focus on scenarios where
the upper $3\times 3$ block of the $M_u$ mass matrix and the remaining
mass matrices are fixed, while the $X_i$, $Y_i$ and $M_{44}$ entries,
related to the new vector-like quark, are free to vary. Furthermore,
we consider separate variations of : (A) only the largest entries,
namely $\{X_2,Y_3,M_{44}\}$, or (B) only (all) the $X_i$
entries. 

To cover the available parameter space while maintaining agreement
with all of the above constraints, we conduct a likelihood analysis
based on Markov chain driven MonteCarlo simulations.  Figures
\ref{fig:Correlations:A} and \ref{fig:Correlations:B} illustrate the
results of such analyses.
In Fig. \ref{fig:Correlations:Aa}, the correlation among the values of
$|V_{ub}|$ and the mass of the new quark $m_T$ is shown for scenario
(A): in that case, accommodating larger $m_T$ values comes at the
price of increasingly larger values of $|V_{ub}|$. To further
illustrate the situation, Fig. \ref{fig:Correlations:Ab} displays the
correlation among $|V_{ub}|$ and $|V_{cb}|$ for separate ranges of
$m_T$ values, $m_T\in[1.00;1.02]$ TeV, $m_T\in[1.10;1.12]$ TeV and
$m_T\in[1.20;1.22]$ TeV. In addition to the features seen in
Fig. \ref{fig:Correlations:Aa}, Fig. \ref{fig:Correlations:Ab} shows
an additional (milder) correlation among $|V_{cb}|$ and $m_T$: larger
masses prefer smaller $|V_{cb}|$ values.
Figure \ref{fig:Correlations:B} corresponds instead to scenario
(B). In this case the correlation trends are reversed with respect to
scenario (A): while $|V_{ub}|$ is almost uncorrelated with $m_T$,
$|V_{cb}|$ tends to be larger for increasing T-masses. It is also to
be noticed that the range of $m_T$ values is much more limited than in
scenario (A).

Finally, we briefly comment on the issue of width of the $750$ GeV
resonance. The first thing to notice is that with the current low
statistics, the estimates for the decay width are very poor. This is
reflected in the fact that, while the ATLAS experiment prefers a broad
decay width of around $45$ GeV, the CMS data suggest a decay width of
few GeV. Such uncertain decay width estimates are likely to change
significantly in the next run, if the anomaly survives.

In our model, since the $\zeta\to TT$ decay is not kinematically
allowed, the decay width of $\zeta$ is a priori narrow.  Even though
$\zeta \to t \bar{t}$ decay is mixing suppressed, it can contribute to
the total width from a few hundred MeVs up to 10 GeVs.
Partial widths to lighter fermions are smaller as well as that for the
$\zeta \to hh$ decay (where h is the \sm Higgs boson) which is
constrained by the LHC Run1 data.
If in future runs the ATLAS experiments confirm that a broad resonance
persists this would imply that a significant novel decay of $\zeta$ is
at work.



This work was supported by MINECO grants FPA2014-58183-P, Multidark
CSD2009-00064 and the PROMETEOII/2014/084 grant from Generalitat
Valenciana. M.N. acknowledges financial support from the
PROMETEOII/2013/017 grant from Generalitat Valenciana.\ RS will like
to thank T. Modak and S. Sadhukhan for useful discussion and
suggestions. The numerical computation was done using MadGraph5aMC@NLO
\cite{Alwall:2014hca} with NN23LO1 PDF set \cite{Ball:2013hta}.

%

%
%
%


\begin{thebibliography}{10}
\providecommand{\url}[1]{\texttt{#1}}
\providecommand{\urlprefix}{URL }
\providecommand{\eprint}[2][]{\url{#2}}

\bibitem{atlas750}
\emph{{Search for resonances decaying to photon pairs in 3.2 fb$^{-1}$ of $pp$
  collisions at $\sqrt{s}$ = 13 TeV with the ATLAS detector}}, Technical Report
  ATLAS-CONF-2015-081, CERN, Geneva (2015),
  \urlprefix\url{http://cds.cern.ch/record/2114853}.

\bibitem{cms750}
C.~Collaboration, \emph{{Search for new physics in high mass diphoton events in
  proton-proton collisions at $\sqrt{s} = 13$ TeV}}  (2015) CMS-PAS-EXO-15-004,
  \urlprefix\url{https://cds.cern.ch/record/2114808}.

\bibitem{Gatto:1968ss}
R.~Gatto, G.~Sartori and M.~Tonin, \emph{{Weak Selfmasses, Cabibbo Angle, and
  Broken SU(2) x SU(2)}}, Phys. Lett. \textbf{B28} (1968) 128--130.

\bibitem{Forero:2014bxa}
D.~Forero, M.~Tortola and J.~Valle, \emph{{Neutrino oscillations refitted}},
  \MYhref[journalLinks]{http://dx.doi.org/10.1103/PhysRevD.90.093006}{Phys.Rev.
  }\MYhref[journalLinks]{http://dx.doi.org/10.1103/PhysRevD.90.093006}{\textbf{D90}
  (2014) 9 093006},
  \MYhref[eprintLinks]{http://arxiv.org/abs/1405.7540}{{\ttfamily
  arXiv:1405.7540 [hep-ph]}}.

\bibitem{Morisi:2013eca}
S.~Morisi et~al., \emph{{Quark-Lepton Mass Relation and CKM mixing in an A4
  Extension of the Minimal Supersymmetric Standard Model}},
  \MYhref[journalLinks]{http://dx.doi.org/10.1103/PhysRevD.88.036001}{Phys.Rev.
  }\MYhref[journalLinks]{http://dx.doi.org/10.1103/PhysRevD.88.036001}{\textbf{D88}
  (2013) 036001},
  \MYhref[eprintLinks]{http://arxiv.org/abs/1303.4394}{{\ttfamily
  arXiv:1303.4394 [hep-ph]}}.

\bibitem{Morisi:2011pt}
S.~Morisi, E.~Peinado, Y.~Shimizu and J.~W.~F. Valle, \emph{{Relating quarks
  and leptons without grand-unification}}, Phys.Rev. \textbf{D84} (2011)
  036003, \MYhref[eprintLinks]{http://arxiv.org/abs/1104.1633}{{\ttfamily
  arXiv:1104.1633 [hep-ph]}}.

\bibitem{King:2013hj}
S.~King, S.~Morisi, E.~Peinado and J.~W.~F. Valle, \emph{{Quark-Lepton Mass
  Relation in a Realistic A4 Extension of the Standard Model}}, Phys. Lett. B
  \textbf{724} (2013) 68--72,
  \MYhref[eprintLinks]{http://arxiv.org/abs/1301.7065}{{\ttfamily
  arXiv:1301.7065 [hep-ph]}}.

\bibitem{Bonilla:2014xla}
C.~Bonilla, S.~Morisi, E.~Peinado and J.~W.~F. Valle, \emph{{Relating quarks
  and leptons with the $T_7$ flavour group}},
  \MYhref[journalLinks]{http://dx.doi.org/10.1016/j.physletb.2015.01.017}{Phys.
  Lett.
  }\MYhref[journalLinks]{http://dx.doi.org/10.1016/j.physletb.2015.01.017}{\textbf{B742}
  (2015) 99--106},
  \MYhref[eprintLinks]{http://arxiv.org/abs/1411.4883}{{\ttfamily
  arXiv:1411.4883 [hep-ph]}}.

\bibitem{Morisi:2012fg}
S.~Morisi and J.~W.~F. Valle, \emph{{Neutrino masses and mixing: a flavour
  symmetry roadmap}}, Fortsch.Phys. \textbf{61} (2013) 466--492,
  \MYhref[eprintLinks]{http://arxiv.org/abs/1206.6678}{{\ttfamily
  arXiv:1206.6678 [hep-ph]}}.

\bibitem{King:2014nza}
S.~F. King et~al., \emph{{Neutrino Mass and Mixing: from Theory to
  Experiment}},
  \MYhref[journalLinks]{http://dx.doi.org/10.1088/1367-2630/16/4/045018}{New
  J.Phys.
  }\MYhref[journalLinks]{http://dx.doi.org/10.1088/1367-2630/16/4/045018}{\textbf{16}
  (2014) 045018},
  \MYhref[eprintLinks]{http://arxiv.org/abs/1402.4271}{{\ttfamily
  arXiv:1402.4271 [hep-ph]}}.

\bibitem{Staub:2016dxq}
F.~Staub et~al., \emph{{Precision tools and models to narrow in on the 750 GeV
  diphoton resonance}}  (2016),
  \MYhref[eprintLinks]{http://arxiv.org/abs/1602.05581}{{\ttfamily
  arXiv:1602.05581 [hep-ph]}}.

\bibitem{Aad:2014fha}
G.~Aad et~al. (ATLAS), \emph{{Search for new resonances in $W\gamma$ and
  $Z\gamma$ final states in $pp$ collisions at $\sqrt s=8$ TeV with the ATLAS
  detector}},
  \MYhref[journalLinks]{http://dx.doi.org/10.1016/j.physletb.2014.10.002}{Phys.
  Lett.
  }\MYhref[journalLinks]{http://dx.doi.org/10.1016/j.physletb.2014.10.002}{\textbf{B738}
  (2014) 428--447},
  \MYhref[eprintLinks]{http://arxiv.org/abs/1407.8150}{{\ttfamily
  arXiv:1407.8150 [hep-ex]}}.

\bibitem{Aad:2015kna}
G.~Aad et~al. (ATLAS), \emph{{Search for an additional, heavy Higgs boson in
  the $H\rightarrow ZZ$ decay channel at $\sqrt{s} = 8\;\text{ TeV }$ in $pp$
  collision data with the ATLAS detector}},
  \MYhref[journalLinks]{http://dx.doi.org/10.1140/epjc/s10052-015-3820-z}{Eur.
  Phys. J.
  }\MYhref[journalLinks]{http://dx.doi.org/10.1140/epjc/s10052-015-3820-z}{\textbf{C76}
  (2016) 1 45},
  \MYhref[eprintLinks]{http://arxiv.org/abs/1507.05930}{{\ttfamily
  arXiv:1507.05930 [hep-ex]}}.

\bibitem{Aad:2015mna}
G.~Aad et~al. (ATLAS), \emph{{Search for high-mass diphoton resonances in $pp$
  collisions at $\sqrt{s}=8$ TeV with the ATLAS detector}},
  \MYhref[journalLinks]{http://dx.doi.org/10.1103/PhysRevD.92.032004}{Phys.
  Rev.
  }\MYhref[journalLinks]{http://dx.doi.org/10.1103/PhysRevD.92.032004}{\textbf{D92}
  (2015) 3 032004},
  \MYhref[eprintLinks]{http://arxiv.org/abs/1504.05511}{{\ttfamily
  arXiv:1504.05511 [hep-ex]}}.

\bibitem{CMS-PAS-HIG-14-006}
\emph{{Search for new resonances in the diphoton final state in the range
  between 150 and 850 GeV in pp collisions at $\sqrt{s} = 8~\mathrm{TeV}$}},
  Technical Report CMS-PAS-HIG-14-006, CERN, Geneva (2014),
  \urlprefix\url{http://cds.cern.ch/record/1714076}.

\bibitem{Agashe:2014kda}
K.~Olive et~al. (Particle Data Group), \emph{{Review of Particle Physics}},
  \MYhref[journalLinks]{http://dx.doi.org/10.1088/1674-1137/38/9/090001}{Chin.Phys.
  }\MYhref[journalLinks]{http://dx.doi.org/10.1088/1674-1137/38/9/090001}{\textbf{C38}
  (2014) 090001}.

\bibitem{Botella:2008qm}
F.~J. Botella, G.~C. Branco and M.~Nebot, \emph{{Small violations of unitarity,
  the phase in $B^0_s - \bar{B}^O_s$ and visible $t \to cZ$ decays at the
  LHC}}, Phys.Rev. \textbf{D79} (2009) 096009,
  \MYhref[eprintLinks]{http://arxiv.org/abs/0805.3995}{{\ttfamily
  arXiv:0805.3995 [hep-ph]}}.

\bibitem{Botella:2012ju}
F.~J. Botella, G.~C. Branco and M.~Nebot, \emph{{The Hunt for New Physics in
  the Flavour Sector with up vector-like quarks}},
  \MYhref[journalLinks]{http://dx.doi.org/10.1007/JHEP12(2012)040}{JHEP
  }\MYhref[journalLinks]{http://dx.doi.org/10.1007/JHEP12(2012)040}{\textbf{12}
  (2012) 040}, \MYhref[eprintLinks]{http://arxiv.org/abs/1207.4440}{{\ttfamily
  arXiv:1207.4440 [hep-ph]}}.

\bibitem{Aad:2016qpo}
G.~Aad et~al. (ATLAS), \emph{{Search for single production of vector-like
  quarks decaying into $Wb$ in $pp$ collisions at $\sqrt{s} =$ 8 TeV with the
  ATLAS detector}}  (2016),
  \MYhref[eprintLinks]{http://arxiv.org/abs/1602.05606}{{\ttfamily
  arXiv:1602.05606 [hep-ex]}}.

\bibitem{Alwall:2014hca}
J.~Alwall et~al., \emph{{The automated computation of tree-level and
  next-to-leading order differential cross sections, and their matching to
  parton shower simulations}},
  \MYhref[journalLinks]{http://dx.doi.org/10.1007/JHEP07(2014)079}{JHEP
  }\MYhref[journalLinks]{http://dx.doi.org/10.1007/JHEP07(2014)079}{\textbf{07}
  (2014) 079}, \MYhref[eprintLinks]{http://arxiv.org/abs/1405.0301}{{\ttfamily
  arXiv:1405.0301 [hep-ph]}}.

\bibitem{Ball:2013hta}
R.~D. Ball et~al. (NNPDF), \emph{{Parton distributions with QED corrections}},
  \MYhref[journalLinks]{http://dx.doi.org/10.1016/j.nuclphysb.2013.10.010}{Nucl.
  Phys.
  }\MYhref[journalLinks]{http://dx.doi.org/10.1016/j.nuclphysb.2013.10.010}{\textbf{B877}
  (2013) 290--320},
  \MYhref[eprintLinks]{http://arxiv.org/abs/1308.0598}{{\ttfamily
  arXiv:1308.0598 [hep-ph]}}.

\end{thebibliography}

%
%
%
%

\end{document}